\documentclass[aps,prl,twocolumn,showpacs]{revtex4}
\usepackage{graphicx}
\usepackage{amssymb}
\usepackage{amsmath}

\begin{document}

\title{Acoustic Emission from Paper Fracture}

\author{L.I. Salminen$^1$, A.I. Tolvanen$^1$, and M.J. Alava$^1$}

\affiliation{$^1$Helsinki University of Technology, Laboratory of Physics,\\ 
P.O.Box 1100, FIN-02015 HUT, Finland }

\begin{abstract}
We report tensile failure experiments on paper sheets.
The acoustic emission energy and the waiting times between 
acoustic events follow power-law distributions.
This remains true while the strain rate 
is varied by more than two orders of magnitude.
The energy statistics has the exponent $\beta \sim 1.25 \pm 0.10$ and
the waiting times the exponent $\tau \sim 1.0 \pm 0.1$,
in particular for the energy roughly independent of the strain rate. 
These results do not compare well with fracture models,
for (brittle) disordered media, which as such exhibit criticality.
One reason may be residual stresses,
neglected in most theories.

\end{abstract}

\maketitle
\date{\today}

A simple, common-day statistical physics experiment
is to tear a piece of paper into two parts. During
this noise is heard \cite{sethrev}, evidently originating 
from damage to the paper or the propagation of a crack.
It is easy to make the 
fracture surfaces ``rough'' as told by the naked eye without
any more sophisticated analysis, and the experience
is that it is difficult to do the tearing so that the
cut is at all close to straight. Meanwhile, one can confirm the existence
of disorder in the ``sample'' by just looking through the sheet
against any reasonable light source. The paper looks cloudy,
which comes to a large degree from the local fluctuations of areal mass
density. This tells that elastic and fracture properties vary locally.
Such a test poses the related questions: how does a crack develop in an
inhomogeneous material, and what kind of properties does the
acoustic emission have?

In fracture there is  evidence of scaling properties familiar from statistical
physics. The roughness of cracks, as measured by e.g.~the
root mean square height fluctuations vs. length scale,
can often be described by self-affine fractal scaling
\cite{Mandel,Bourev,DNBC}. 
Another such quantity is the distribution of 
strength, and its average versus sample size. 
In particular for (quasi)-brittle materials, 
the average strength can follow scaling laws, if the sample size is varied, 
which derive from the presence of disorder in the material 
\cite{Dux,Curtin,Korte}.

Investigating such questions combines materials
science and statistical mechanics. Two fundamental problems are, 
what is the role of disorder in fracture? And, how do
cracks develop and interact in real materials? These join forces
in the failure of a notched three-dimensional sample,
visualized by the advancement of an one-dimensional crack front or line in
the presence of disorder \cite{HaH95,Fisher}. 
It has an equation of motion, in an environment with 
varying properties as local strength and elastic modulus.
The trail left by the line-like front is the two-dimensional crack
surface, which can become rough (self-affine)
if the propagating line-like object develops critical fluctuations. However, it is difficult to propose
such a model that would match with the experimental data.
For one, simple models tell that the disorder and microcracks 
interact in crack dynamics, so that sometimes it is not correct
to consider an isolated object like the front or a crack tip. 
From the materials research viewpoint the dynamics is interesting
since it deals with the engineering quantities
of strength and toughness, or the resistance to intrinsic flaws. 
For instance in fiber and two-phase ceramic composites
one can control these two by varying the mixing and constituents.

We study fracture in ordinary paper, as a two-dimensional
disordered material, via acoustic
emission (AE) analysis. The release of acoustic 
energy is related in paper to irreversible deformation, microcracks,
and, perhaps, to plasticity. A large-scale analogy is
earthquakes. Their energies are described by 
a power-law probability distribution, the Gutenberg-Richter-law
\cite{stars} with an {\em energy exponent} $\beta$.
Cracks in paper may be self-affine
\cite{Kertesz,Rosti},
with a roughness exponent close to that in a scalar two-dimensional
fracture model, random fuse networks (RFN's) 
\cite{Maloy,letteri}. Both these are near
2/3, the one for 
surfaces of minimal energy (\cite{HaH95,letteri,Rosti}).
Paper provides an example
of crack growth in the presence of disorder. 
The probability distributions of the released energy
and the temporal statistics, describing dynamical aspects,
can be compared to models for two-dimensional failure,
and to other work on AE. 

Here the {\em strain rate} $\dot{\epsilon}$ is varied in the tensile test,
to study possibly time-dependent effects in the fracture process.
Our first main conclusion is that in strain-controlled tests
power-laws are found in the statistics, with either no 
or at most a weak dependence on the strain rate. The power-laws
do not follow the predictions of theoretical models outlined
below, as fuse networks or mean-field ones. It is also apparent
that an eventual localization of the crack \cite{Ciliberto,penn}
does not change the microscopic properties of the 
crackling noise, as long as the crack propagation is stable.
This implies that paper failure is not a 
``phase transition'' with a diverging correlation length,
but leaves open the origins of the observed scalings.

To describe the interactions in an elastic medium
with randomness incorporated one starts from mean-field 
stress-sharing. In fiber bundle models (FBM) the applied external
force $F(t)$ is shared by all the $N(t)$ fibers,
and as the (random) failure threshold of the currently weakest
one is reached, the stress $\sigma(t) = F(t)/N(t)$
is instantaneously distributed among the remaining fibers. 
As a sign of criticality
there is a divergent scale, the size of a typical avalanche
which is made up of all the fibers that break down due to a single,
slow-time scale stress-increment. The total avalanche size ($N$) distribution
follows $P(N) \sim N^{-5/2}$ \cite{hansen,sornette}. 
Thus global load sharing 
fracture resembles a second-order phase transition. 

Local stress enhancements can be added to FBM's e.g. by
considering fiber chains in which the stress from 
microcracks, of adjacent 
missing fibers, is transfered to the ones neighboring
the crack. 
A more catastrophic growth results (with an exponent
much larger than 5/2), resembling a
first-order phase transition since the elastic modulus has a finite
drop at $\sigma_c$/$\epsilon_c$ \cite{zhang,hansen,phoenix}. 
A finite-dimensional, but scalar, approximation
is given by random fuse networks \cite{review}.
RFN's reproduce with strong enough disorder many of
the features of mean-field FBM's, in spite of 
including the stress-enhancements
and shielding from other microcracks \cite{Zapperi,Z2}.
The dynamics of crack growth is not understood well  even in this 
simple model, including why the cracks
seem to become self-affine, with a roughness exponent close to 
2/3 \cite{letteri}.
The models' event statistics can be compared with experiments
by considering the energies.
For the RFN, in an event the lost energy is
$E_i \sim \Delta G_i \epsilon_i^2$, where $\Delta G_i \sim N_{broken}$ 
is the change in the ``elastic modulus''
due to the AE event, at $\epsilon_i$. 
Simulations on RFN's for strong disorder that reproduces
the FBM exponent (5/2) for the avalanche sizes yield the
exponent $\beta_{RFN} \sim 1.8$ and the 
FBM value is $\beta_{FBM} \sim 2$ \cite{Rostia}.

Experiments on, mostly, three-dimensional
systems \cite{Ciliberto, Petri2, Lockner, geofyysikoita,penn}
have yielded power-laws for the AE energy release; the typical
exponent $\beta$ is 1 $\dots$ 1.5. One idea
is that approaching final failure resembles
 a phase transition: in the 
Lyon group's stress-controlled experiments indications were seen of a 
critical energy release rate \cite{Ciliberto}. 
This means that the AE energy would diverge
like a power-law in the proximity of a critical point, 
the sample strength $\sigma_c$. 
Another problem is the nature of the
disorder in the material: whether quenched disorder
is able to give such power-law statistics 
for the energy \cite{Petri2,Zapperi}.

Normal newsprint paper samples were tested in the machine direction
\cite{md/cd} on a mode I laboratory testing machine of type MTS 400/M.
Due to the lack of constraints the samples could have out-of-plane
deformations, and none of the three fracture modes (I, II, III) is
excluded on the microscopic level.
The deformation rates $\dot{\epsilon}$ varied between 0.1 \%/min and 100
\%/min.  The AE system consisted of a piezoelectric transducer, a rectifying
amplifier and continuous data-acquisition. The time-resolution of 
the measurements was 10 $\mu$s and the data-acquisition was free of deadtime. 
The stress was measured
simultaneously to AE with a time resolution of 0.01 s.  We made 20 identical
repetitions for statistics.  The 100 by 100 mm samples had initial notches 
(size 15 mm) to achieve stable crack growth. Typical sheet thickness is
about 100 $\mu$m. The fracture statistics are not
affected by such a notch \cite{Rostia}. For these strain rates the
sound velocity timescale is much faster than that implied by
$1/\dot{\epsilon}$. Each individual test contributes, at most,
1000 - 2000 events,
so we can only look at integrated probability distributions and
not in detail at local averages of e.g the acoustic event size vs. 
$\epsilon$.

\begin{figure} 
 \includegraphics[width=7cm
]{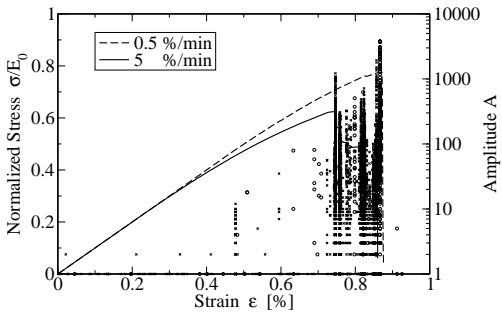}
\caption{The acoustic data series (crosses and circles) and the
the stress-strain curves, for two strain rates. There
is a difference in the post-stress
maximum part of the stress-strain curve, which
is more important
for  $\dot{\epsilon}$ 0.5 [\%/min] (solid line and crosses)
than 5 [\%/min] (dashed line and circles).}
\label{fig1}
\end{figure}

Fig.~(\ref{fig1}) shows an example of two tests
under strain-control. Stress-strain 
curves have typically three parts: pre-failure (almost) linearly
elastic one, the regime close to the maximum stress where the
the final crack starts to propagate
or is formed, and a tail that arises due to the cohesive
properties of paper which allow for stable crack propagation.
The faster the strain rate the less is the role of the tail.
For the smallest strain rate most of the AE originates from
tail (more than 90 \%)
while for the highest the situation is the opposite.
Quantities of interest are
the statistical properties before the maximum, after it,
and the integrated totals, in particular the energy distributions.
The time series of events allows to make qualitative observations
of the correlations between subsequent events and to draw conclusions
about the event properties as such. For this rather brittle paper 
grade, and the strain rates used,
the elastic modulus is independent of 
$\dot{\epsilon}$,
and by AE we are able to detect a constant fraction of the elastic
energy \cite{Rostia}.

\begin{figure}
 \includegraphics[width=7cm
]{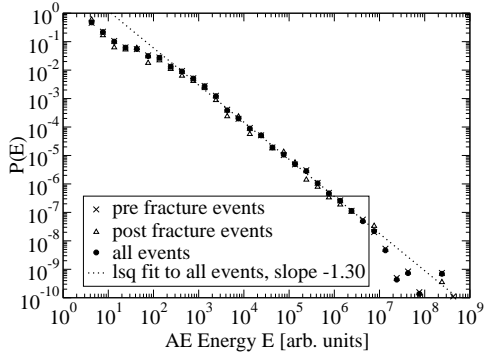}
\caption{The three (post/pre/total)
energy distributions, $\dot{\epsilon}$ 1.0 [\%/min] 
(for which the post/pre -contributions to AE energy
are roughly the same).}
\label{fig2a}
\end{figure}

Fig.~(\ref{fig2a}) shows the scaling of the
energy for a fixed strain rate. The behavior
is power-law -like, with
several orders of magnitude of scaling.
The same exponent fits all three different
cases: the pre and post maximum stress cases,
and the sum distribution. If the strain rate
is varied the same conclusion holds and the exponent
only fluctuates at random \cite{Rostia}.
We have $\beta = 1.25 \pm 0.10$,
in disagreement 
with the fiber bundle ones and with that from the
fuse networks, though $\beta_{RFN}$ is closer than $\beta_{FBM}$.
The practical implication is, that the material can 
withstand more damage than expected since $\beta < \beta_{models}$.
In paper, the post-maximum events correlate with
the advancement of the final crack. The remaining ligament length/width 
contracts, thus the elastic modulus drops should (assuming a
constant stress state) here be qualitatively
related to AE event energies (see also \cite{balankin}).

In an experiment with a varying strain rate $\dot{\epsilon}$ 
both the event durations and the waiting time between
any two events may (consider Fig.~(\ref{fig1})) depend on
$\dot{\epsilon}$. If the crack dynamics becomes ``fast'' then
events take place on timescales set by the sound velocity.
This establishes a timescale $t_s = \Delta x / v_s$ where $\Delta x$
is the spatial separation of two events and 
the sound velocity $v_s \sim$ 2 $\times$ 10$^3$ m/s. In a sample
of linear size 0.1 m this results in a maximum $t_s \sim 10^{-4} s$.
In the failure of brittle carbon foams the eventual critical crack growth
is dictated by such fast events \cite{penn}.

In quasi-static fracture models the dynamics
of cracks is assumed such that the stress field is equilibrated
infinitely fast during microfracture events, between further 
adiabatic increases of strain or stress. Thus the only time-dependence
of any temporal statistics is in the average time interval, between 
AE events, proportional to $1/\dot{\epsilon}$. For the FBM model
there are no correlations between the waiting times and event
size or durations, except for the trend that the average waiting
time decreases as $\sigma_c$ is approached. It is 'critical' and
thus the energy release rate follows a power-law close to
$\sigma_c$.  
Waiting time results do not generally exist for more complicated
models \cite{herrmann}.
One can compare to the experimental signatures
of the intervals between AE events and the integrated energy
release rate integral $\int d\epsilon E_{event}$.
As noted above there is some evidence - from stress-controlled
experiments - for a critical behavior for this quantity  \cite{Ciliberto}.

\begin{figure}

 \includegraphics[width=7cm
]{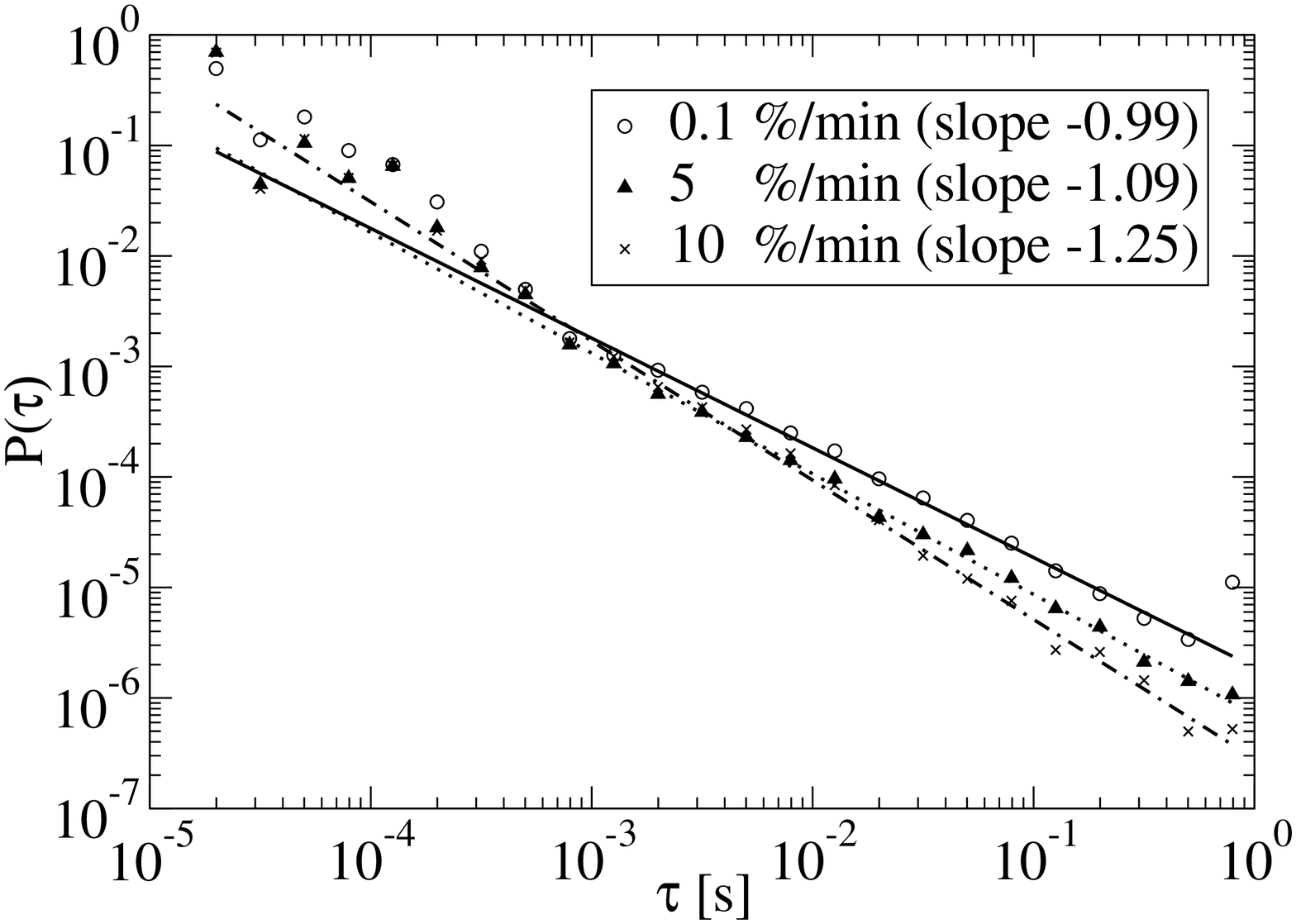}
\caption{Plots of the event interval $\tau$ (waiting time)
distributions, for three strain rates.}
\label{fig3b}
\end{figure}

\begin{figure}
 \includegraphics[width=7cm,angle=-0
]{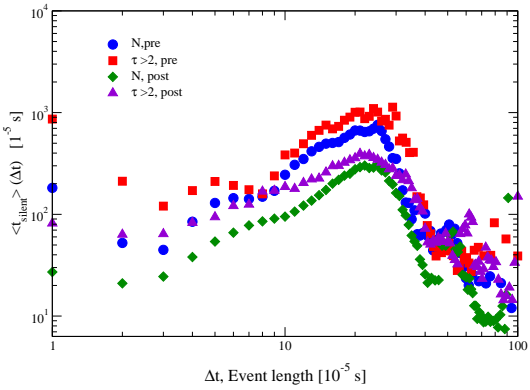}

\caption{The waiting time as a function of
event length $\Delta t$,
$\dot{\epsilon}$ 1.0 [\%/min].
}
\label{fig4}
\end{figure}

Fig.~(\ref{fig3b}) demonstrates waiting time 
$\tau$ distributions for different strain rates. 
There is a clear power-law, whose exponent ($\tau$) 
remains roughly the same 
for all the strain rates \cite{cut-off}. 
Importantly, this is true for the post-events regardless 
of the origin (before/after $\sigma_c$) of the majority
of the AE energy.
For the pre-events there 
might be some evidence of the exponent increasing with
$\dot{\epsilon}$.
In the time-series of events,
those with long durations are separated, on the average
by shorter intervals from the neighboring events
before/after. Fig.~(\ref{fig4})
depicts the waiting times prior to an event with
two different data analysis methods for distinguishing
between possibly overlapping events \cite{Rostia}.
The interval separations are similar for both the post/pre-phases, 
implying that the microscopic failure dynamics does not differ,
and that $\sigma_c$ or $\epsilon_c$ can not be inferred from
the event characteristics. It may be so
that events which are relatively long are precluded by
longer waiting times. The durations of the events $\delta t$
and the sizes are roughly power-law related as
$\langle E \rangle \sim (\delta t)^3$.

The integral $\int d\epsilon E_{event}$ demonstrates a rapid
exponential growth above a typical strain of $\epsilon \sim 0.5$
\%. This originates from increasing event sizes, not from an
increasing density vs. strain. For $\epsilon > 0.6$ \% the samples
start to fail. In the regime where the exponential growth
takes place the samples develop plastic, irreversible
strain $\epsilon_{pl}$, with a roughly exponential dependence on strain 
\cite{Rostia}.
This does not imply that the AE measures plastic
deformation work, mostly, since the rate of increase
of $\epsilon_{pl}$ is much less than that of the energy
integral. 
The exponential strain-dependence of
the AE implies a typical lengthscale (as should be true
for the development of plasticity), which in turn should be related 
to crack localization.

Concluding, failure of ordinary paper shows several
features associated with critical phenomena. 
Our take on the experiment is that it
show that {\em i)} there is no clear sign of a ``critical
point'', or a phase transition, in spite of the fact that
the material is close to linearly elastic. In particular,
while the event intervals and the event energies follow
power-law -like statistics, not all the quantities do so.
Also, {\em ii)} the temporal
behavior hints of complicated time-dependent phenomena
not directly related to the fast relaxation of stress. A possible
candidate is the viscoelastic nature of the wood fibers in
paper, but note that the 
macroscopic stress-strain behavior remains almost linearly
elastic while AE events already occur. We suspect the
gradual release of internal stresses plays a role,
perhaps due to frictional pull-out of fibers from the
network.
Finally, {\em iii)}
the power-laws as obtained are off those predicted by
simple fracture models. It remains to be seen whether these
models can be tailored closer to such tensile experiments \cite{visco}.
One suggestion is that
the dynamics of energy release during the events follows
a different course from the
model rules, in particular finite-rate dynamics allows
for stress overshoots \cite{fiprl,recent}. 
The relation of the acoustic emission to why cracks get rough
may be indirect. The latter could relate to the development
of pre-failure plastic deformation. 

We acknowledge support by the Academy of Finland's Center
of Excellence program, and discussions with K. Niskanen,
J. Weiss, and S. Zapperi.

\begin{figure}
 \includegraphics[width=7cm
]{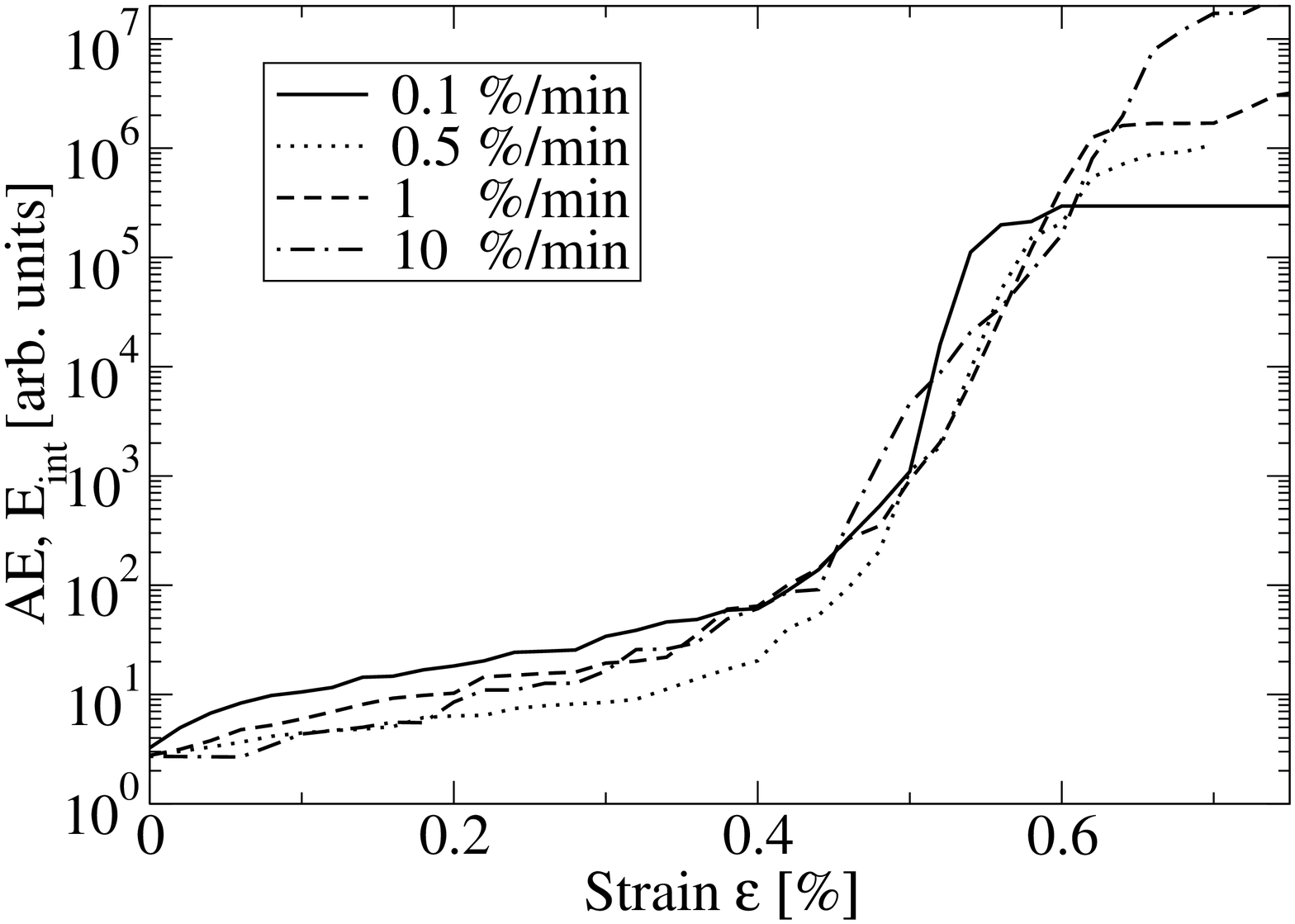}
\caption{Increase of the integrated AE energy vs.
strain $\epsilon$.}
\label{fig2}
\end{figure}

\begin{references}
\bibitem{sethrev} For ``crackling'' noise, see
J.P. Sethna, K.A. Dahmen, and C.R. Myers, 
	Nature (London) {\bf 410}, 242 (2001). One can also
just crumble paper and listen; P.A. Houle and J.P. Sethna,
Pys. Rev. {\bf E54}, 278 (1996).
\bibitem{Mandel} B.~B.~Mandelbrot, D.~E.~Passoja, and A.~J.~Paullay,
 Nature (London) {\bf 308}, 721 (1984).
\bibitem{Bourev} E.\ Bouchaud, J.\ Phys.\ Cond.\ Mat.\ {\bf 9}, 4319 
(1997).
\bibitem{DNBC} P.~Daguier {\em et al.},
Phys. Rev. Lett. {\bf 78}, 1062 (1997).
\bibitem{Dux}
P.\ M.\ Duxbury, P.\ D.\ Beale, and P.\ L.\ Leath, 
Phys.\ Rev.\ Lett.\ {\bf 57}, 1052 (1986).
\bibitem{Curtin}  W.A. Curtin,
	Phys. Rev. Lett. {\bf 80}, 1445 (1998).
\bibitem{Korte} M. Korteoja {\em et al.}, Mat. Sci. Eng. {\bf A240},
173 (1999).
\bibitem{HaH95} T.\ Halpin-Healy and Y.-C.\ Zhang, Phys. Rep. {\bf 254},
215 (1995).
\bibitem{Fisher} S. Ramanathan, D. Erta\'s, and D.S. Fisher,
Phys. Rev. Lett. {\bf 79}, 873 (1997).
\bibitem{stars} V.G. Kossobokov, V.I. Keilis-Borok, and B. Cheng,
Phys. Rev. {\bf E61}, 3529 (2000).
\bibitem{Kertesz} J.\ Kert{\'e}sz, V.\ K.\ Horvath, and F.\ Weber, Fractals
{\bf 1}, 67 (1993).
\bibitem{Rosti}
J. Rosti {\em et al.},
Eur. Phys. J. {\bf B19}, 259 (2001).
\bibitem{Maloy} T. Engoy {\em et al.},
Phys. Rev. Lett. {\bf 73}, 834 (1994).
\bibitem{letteri} V.\ I.\ R\"ais\"anen {\em et al.}, 
Phys.\ Rev.\ Lett.\ {\bf 80}, 329 (1998);
E.T. Sepp\"al\"a, V.I. R\"ais\"anen, and M.J. Alava,
Phys. Rev. {\bf E61}, 6312 (2000).
\bibitem{Ciliberto}
A. Guarino, A. Garcimartin, and S. Ciliberto,
Eur. Phys. J. B {\bf 6}, 13 (1998);
A. Garcimartin {\em et al.}, 
Phys. Rev. Lett {\bf 79}, 3202 (1997).
\bibitem{penn}
L.C. Krysac and J.D. Maynard, 
	Phys. Rev. Lett. {\bf 81}, 4428 (1998).
\bibitem{hansen}
M. Kloster, A. Hansen, and P.C. Hemmer, 
	Phys. Rev. {\bf E56}, 2615 (1997).
\bibitem{sornette}
D. Sornette, J. Phys A. {\bf 22}, L249 (1989).
\bibitem{zhang}
Shu-dong Zhang 
Phys. Rev. {\bf E59}, 1589 (1999)
\bibitem{phoenix} W.I. Newman and S.L. Phoenix, 
Phys. Rev. {\bf E63}, 021507 (2001).
\bibitem{review} Chapters 4-7 in {\it Statistical models for the
fracture of disordered media}, ed. H.\ J.\ Herrmann and S.\ Roux,
(North-Holland, Amsterdam, 1990).
\bibitem{Zapperi} S.~Zapperi {\em et al.},
Phys. Rev. Lett. {\bf 78}, 1408 (1997). 
\bibitem{Z2} The same may be true for vectorial
models: S.~Zapperi {\em et al.},
Phys. Rev. {\bf E59}, 5049 (1999).
\bibitem{Rostia}
L. Salminen {\em et al.}, in preparation.
\bibitem{Petri2}
A. Petri {\em et al.}, 
	Phys. Rev. Lett {\bf 73}, 3423 (1994).
\bibitem{Lockner}
D.A. Lockner {\em et al.}, 
	Nature {\bf 350}, 39 (1991).
\bibitem{geofyysikoita}
J. Weiss, J.R. Grasso,  and P. Martin, 
Proc. 6th Int. Conf. on AE/MS in Geol. Struct. \&
Mat., 1996, 583-595, Trans Tech Publications,
(Clausthal-Zellerfeld).
\bibitem{Petri}
G. Caldarelli, F.D. Di Tolla, and A. Petri, 
	Phys. Rev. Lett. {\bf 77}, 2503 (1996).
\bibitem{md/cd} Industrial paper has two main orientations,
machine and crossmachine-directions. It is much less
ductile in the former. See eg. K.J. Niskanen (ed.), {\it Paper Physics}, 
(Fapet, Helsinki, 1998). 
\bibitem{balankin}
A.S. Balankin, O. Susarrey, and A. Bravo,
	Phys. Rev. {\bf E64}, 066131 (2001). 
\bibitem{herrmann} An exception is F. Tzschichholz and H.J. Herrmann,
Phys. Rev. {\bf E51}, 1961 (1995).
\bibitem{cut-off} Notice that the largest values of $\tau$
are still much smaller than the maximal timescale, given by
$\epsilon_c /\dot{\epsilon}$.
\bibitem{visco}
Eg. fiber bundle models can be made time-dependent, see:
R.C. Hidalgo, F. Kun, and H.J. Herrmann,
	Phys. Rev. {\bf E65}, 032502 (2002). 
\bibitem{fiprl} J.M. Schwarz and D.S. Fisher,
Phys.\ Rev.\ Lett.\ {\bf 87}, 096107 (2001).
\bibitem{recent} Adding dynamics to a model does not necessarily change
$\beta$, M. Minozzi {\em et al.}, cond-mat/0207433.
\end{references}
\end{document}